\documentclass[fleqn,twoside]{article}
\usepackage{espcrc2}
\usepackage{textcomp}

\usepackage[dvips]{graphicx}
\usepackage[figuresright]{rotating}

\newcommand{\AmS}{{\protect\the\textfont2
  A\kern-.1667em\lower.5ex\hbox{M}\kern-.125emS}}

\newcommand{\E}{E_{0}}
\newcommand{\Eeph}{E_{\mathrm{m}}}
\newcommand{\Eth}{E_{\mathrm{th}}}
\newcommand{\rhom}{\rho_{\mu}}
\newcommand{\rhos}{\rho_{\mathrm{s}}}
\newcommand{\sigmean}{\sigma_{\mu}^{\mathrm{mean}}}
\newcommand{\sigphys}{\sigma_{\mu}^{\mathrm{phys}}}
\newcommand{\sigmeth}{\sigma_{\mu}^{\mathrm{meth}}}
\newcommand{\Xmax}{X_{\mathrm{max}}}
\newcommand{\chis}{\chi^{2}}
\newcommand{\Nexp}{N_{\mathrm{exp}}}
\newcommand{\Nteor}{N_{\mathrm{theor}}}

\title{Characteristics of EAS and Primary Particle Mass Composition in the Energy Region
of $10^{17} - 3 \cdot 10^{19}$~eV by Yakutsk Data}

\author{S.P.~Knurenko\address[YKT]{Yu.G.Shafer Institute of Cosmophysical Research and
        Aeronomy, \\
		31 Lenin Ave., 677980 Yakutsk, Russia},
		V.P.~Egorova\addressmark[YKT]{},
		A.A.~Ivanov\addressmark[YKT]{},
		V.A.~Kolosov\addressmark[YKT]{},
		I.T.~Makarov\addressmark[YKT]{},
		Z.E.~Petrov\addressmark[YKT]{},
		I.Ye.~Sleptsov\addressmark[YKT]{},
		G.G.~Struchkov\addressmark[YKT]{}.
		}

\begin{document}

\begin{abstract}
In the work the data of the Yakutsk complex EAS array and their comparison with
calculation in the case of primary nuclei of different chemical elements are presented.
The calculation by QGSJET model have been used interpreting experimental data.
\vspace{1pc}
\end{abstract}

\maketitle

\section{INTRODUCTION}

It is not possible to measure a mass composition of primary cosmic rays (PCR) in the
energy range $10^{15} - 10^{20}$~eV using the direct method. Nothing remains, but to
resort to the indirect methods when for the similar estimates the measurement of different
components of extensive air shower  (EAS) are used. That can be characteristics of
longitudinal or lateral development of shower in the air. And usually it is connected with
the analysis of the most sensitive to the composition of shower components which differ
from each other by the character of formatting and absorbing in the atmosphere, for
example, of the charged particle flux (electron, muon), the flux of \^{C}erenkov or
ionization radiation.

\section{ENERGY TRANSFERRED TO THE ELECTROMAGNETIC EAS COMPONENT}

\begin{figure}
\includegraphics[width=7.5cm]{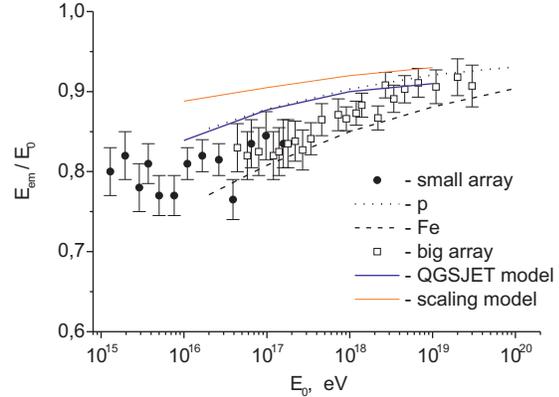}
\caption{A portion of the energy transferred to the electromagnetic EAS component by
\^{C}erenkov light data at the Yakutsk array.}
\label{fig1}
\end{figure}

Fig.\ref{fig1} presents the EAS Yakutsk array experimental data and calculations by the
model with the decelerated and moderate dissipation of the energy into the electromagnetic
EAS component: quasiscaling models (solid line) and QGSJET (dashed line)~\cite{bib1}. From
Fig.\ref{fig1} it is seen both the agreement of experimental data and calculations by
QGSJET model (proton) in the region $\E \ge 3 \cdot 10^{18}$~eV, and disagreement at $\E
\le 3 \cdot 10^{18}$~eV. The scaling model gives a noticeably greater value of $\Eeph /
\E$ in relation to the experimental data that is doubtlessly also connected with the break
of  scaling function in the region of ultra--high energies.

The experimental data in Fig.\ref{fig1} is well approximated by the expression of a form:
\begin{eqnarray}
\frac{\Eeph}{\E}  & = & (0.964 \pm 0.011) - (0.079 \pm \nonumber \\
&& \pm 0.005) \cdot \E^{- (0.147 \pm 0.008)}\mathrm{.}
\label{eq1}
\end{eqnarray}
The relation (\ref{eq1}) is primarily important for the comparison of estimations of $\E$
obtained at the Yakutsk and Fly's Eye arrays~\cite{bib2}.

The calculations in~\cite{bib2} (see Fig.\ref{fig1}) have been carried out by the QGSJET
model in the case of primary proton and iron nucleus. A good agreement of our calculations
in the case of the primary proton is observed. The comparison of experimental data with
calculations for the proton and iron nucleus indicates to the fact that the mass
composition of particles of cosmic radiation in the energy region of $10^{17} -
10^{18}$~eV and above $3 \cdot 10^{18}$~eV must differ. At $\E \ge 3 \cdot 10^{18}$~eV the
mass composition is most likely close to the proton one.

\section{A PORTION OF MUONS WITH $\Eth \ge 1$~GeV}

\begin{figure}
\includegraphics[width=7.5cm]{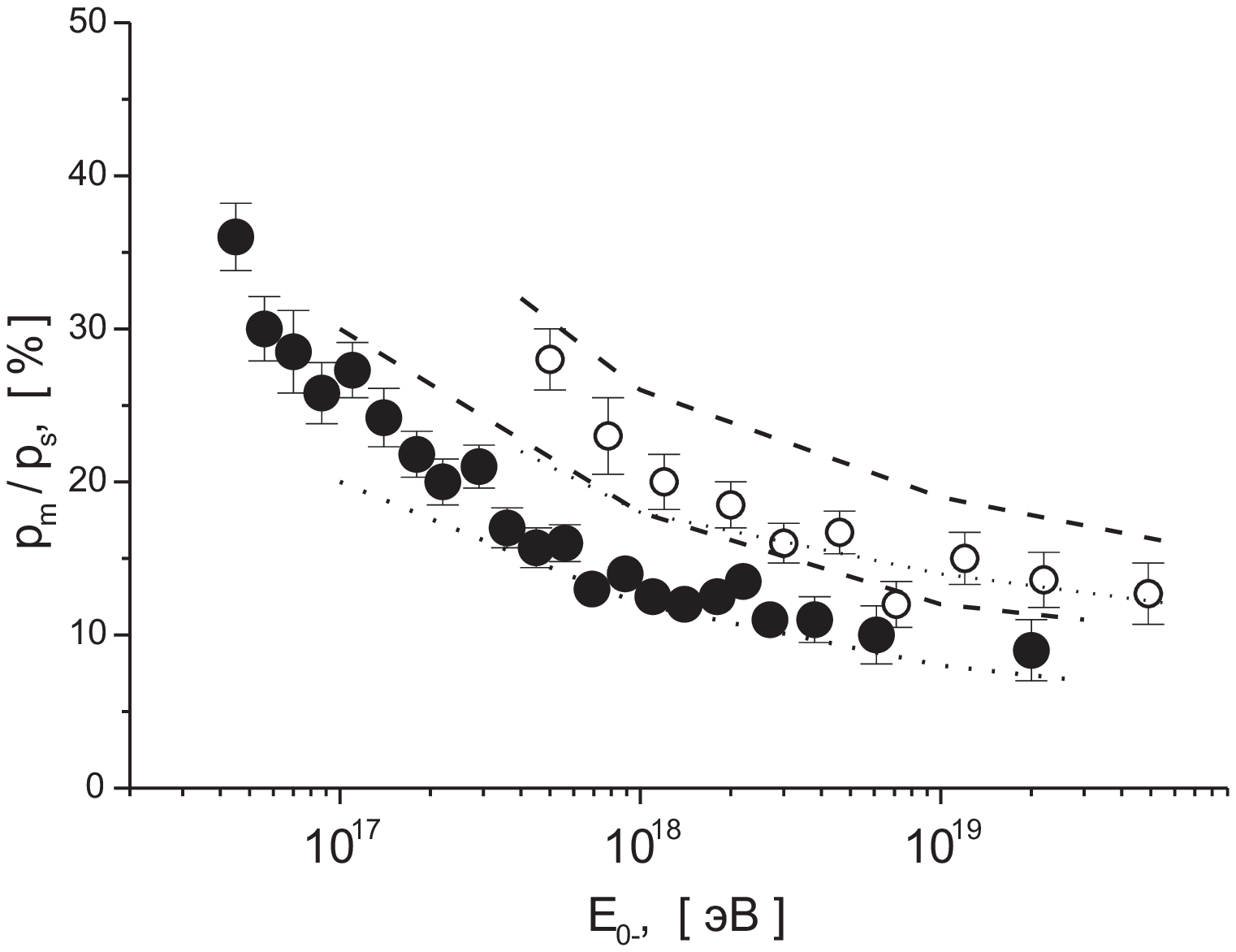}
\includegraphics[width=7.5cm]{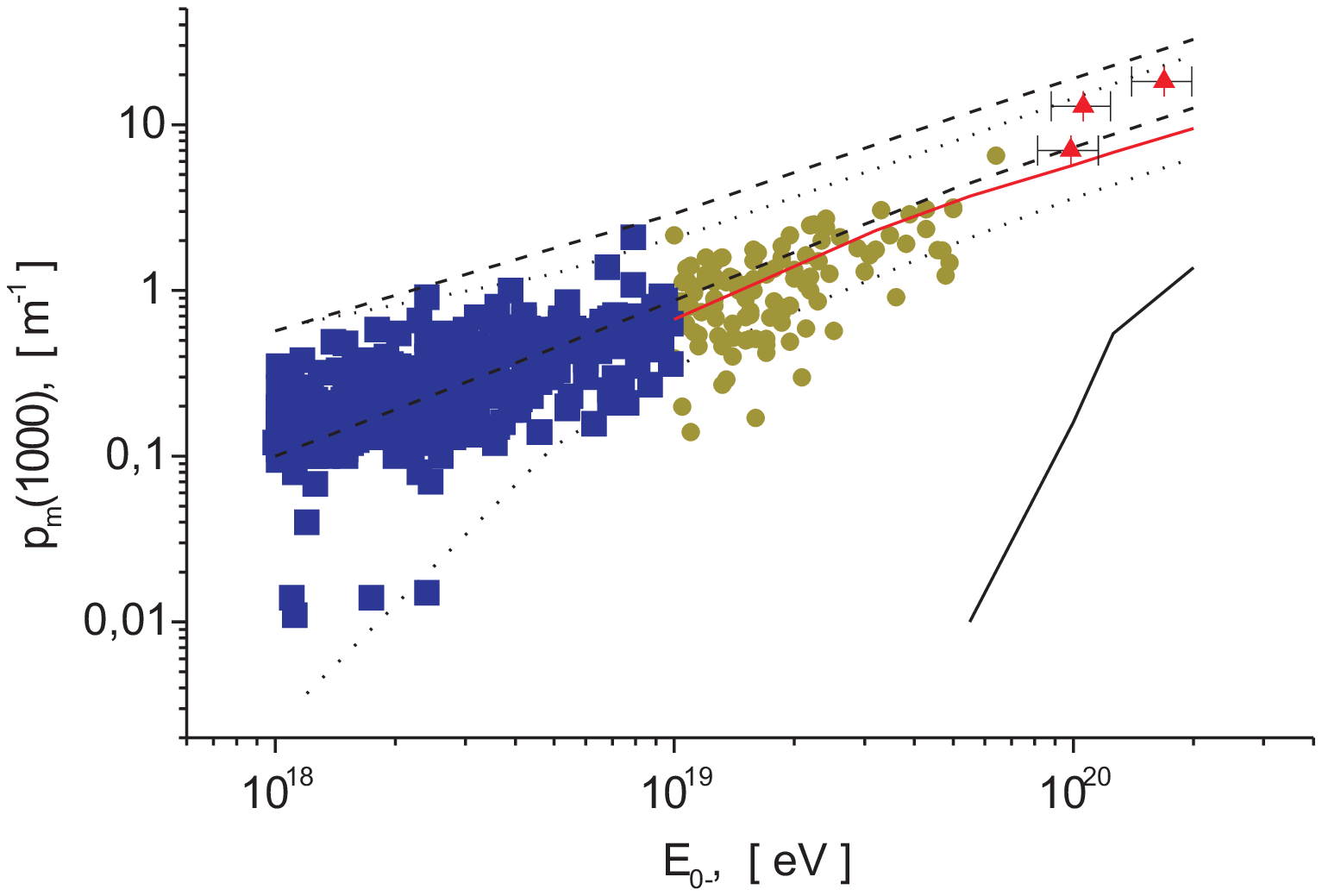}
\caption{a) Portion of the muons with $\Eth \ge 1$~GeV (\%). $\rhom(300) / \rhos(300)
(\bullet)$ and $\rhom(600) / \rhos(600) (\circ)$. b) $\rhom(1000)$ vs. $\E$ relations for
observed events $10^{18} - 10^{19}$~eV (square), $10^{19} - 10^{20}$~eV (points) and
$10^{20} - 10^{21}$~eV (triangles). Expected $\pm 1 \sigma$ bounds for the distributions
are indicated for proton, iron and gamma--ray primary by different curve as in the
legend.}
\label{fig2}
\end{figure}

The showers with $\E \ge 10^{18}$~eV and zenith angles $\theta < 60^{\circ}$ have been
chosen. It is required so that the shower axis is within the array and not less than three
muon detectors (one of them is at a distance of 1000~m from the shower axis) operate
during the shower. The density of muons flux $\rhom(1000)$ is like a median between the
densities adjusted to $\left<R\right> = 1000$~m, according to a mean lateral muon
distribution function. In this case the detectors which have been operated in the shower
(they have given zero indications), i.e. the detector threshold were used in analyze too.
The muon flux densities at have distances of 300 and 600~m from the shower axis have been
calculated in analogous way. Results are in Fig.\ref{fig2}a and \ref{fig2}b.
Calculations~\cite{bib3} have shown that $\rhom(R)$ weekly depends on a zenith angle in
the interval of $\Delta \theta = (0^{\circ} - 60^{\circ})$. So for the comparison of
calculations and experimental data we take a mean angle $\theta = 39^{\circ}$. In order to
take into account both physical fluctuations in measurement of $\rhom(1000)$ and
methodical ones (apparatus errors and accuracy of the locate of the shower axis) we have
considered the judicial interval in one $\sigmean =  \sigphys + \sigmeth$. In
Fig.\ref{fig2}a and \ref{fig2}b the calculation result is shown by
dotted line in the case of primary proton and by a dashed line for the iron nuclei. A
solid line shows the upper and lower limits for the case if EAS with $\E \ge 10^{19}$~eV
would generate by a primary $\gamma$--quantum. The calculation for the primary
$\gamma$--quantum has been taken from~\cite{bib4}. Dots for $\rhom(300)$ and circles for
$\rhom(600)$ the experimental data show in Fig.\ref{fig2}a. In Fig.\ref{fig2}b the showers
in the energy range of $10^{18} - 10^{19}$~eV are shown by squares, the showers with $\E
\ge 10^{19}$~eV are shown by dots and the showers of maximum energy are shown by
triangles. The comparison of experimental data presented in Fig.2a with calculations by
the QGSJET model carried out for the case of the primary proton and iron nuclei confirms
the hypothesis on the fact that the considerable portion of ultimate energy EAS have been
formed by protons. Their portion decreases below the energy $10^{18}$~eV. It is seen from
Fig.2b, the basic mass of points (showers with $\E \ge 10^{19}$~eV) falls into the
interval for the proton. 23 points fall into a zone of superposition of a proton and iron
nuclei and 19 showers from 116 fall into a zone of upper boundary of calculation for the
primary $\gamma$--quantum.  It testifies to the fact that even taking into account there
exists a probability that showers of such energies can be generated by neutral particles
and, in particular, by a primary $\gamma$--quantum. Then it is justified to use the
analysis of directions of arrival of showers with $\E \ge 10^{19}$~eV for the search of
the sources of highest energy cosmic rays. In this case the determination accuracy of
arrival angles of such showers must be not worse than $(0.5^{\circ} - 1.5^{\circ})$.
\begin{figure}
\includegraphics[width=7.5cm]{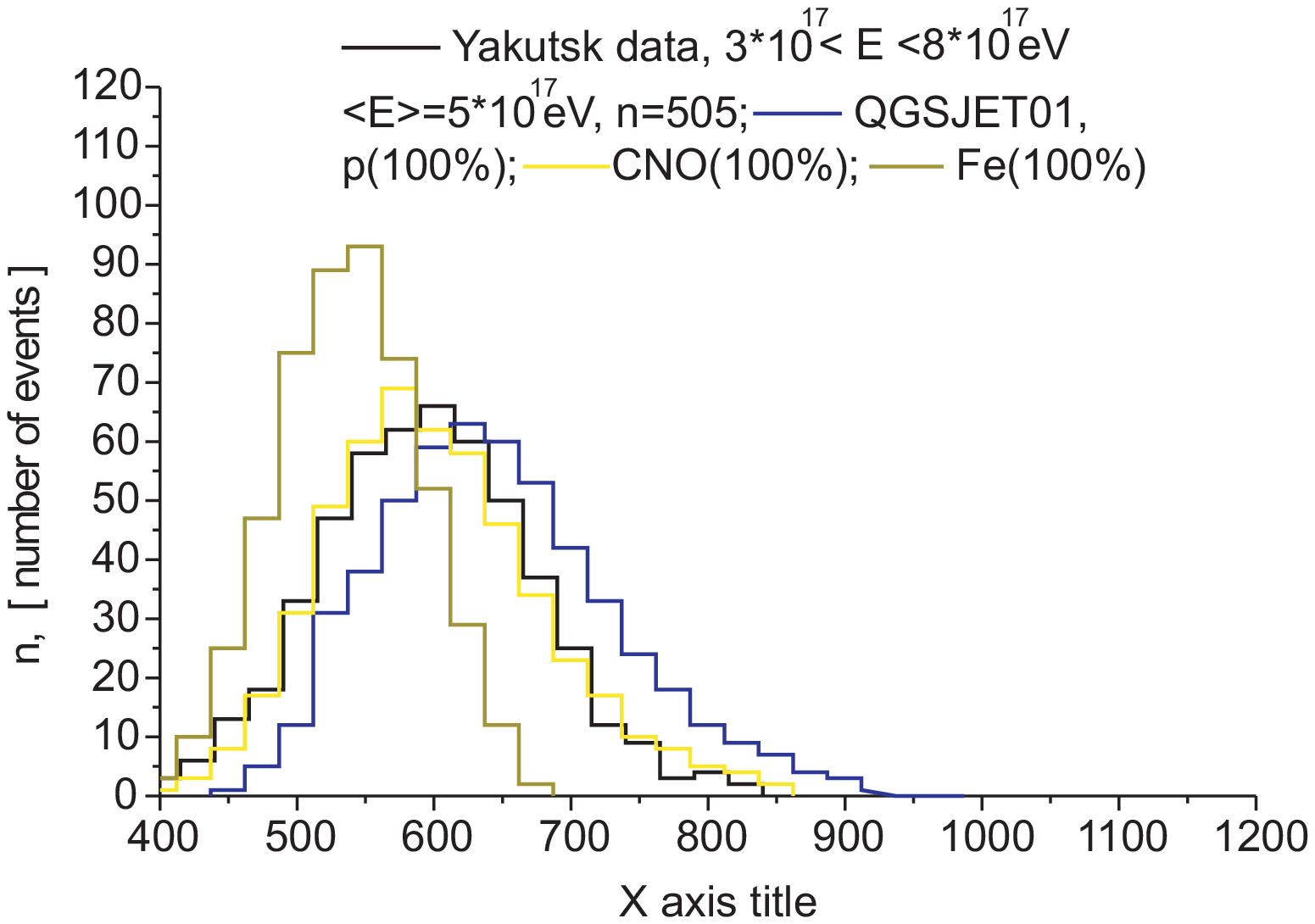}
\includegraphics[width=7.5cm]{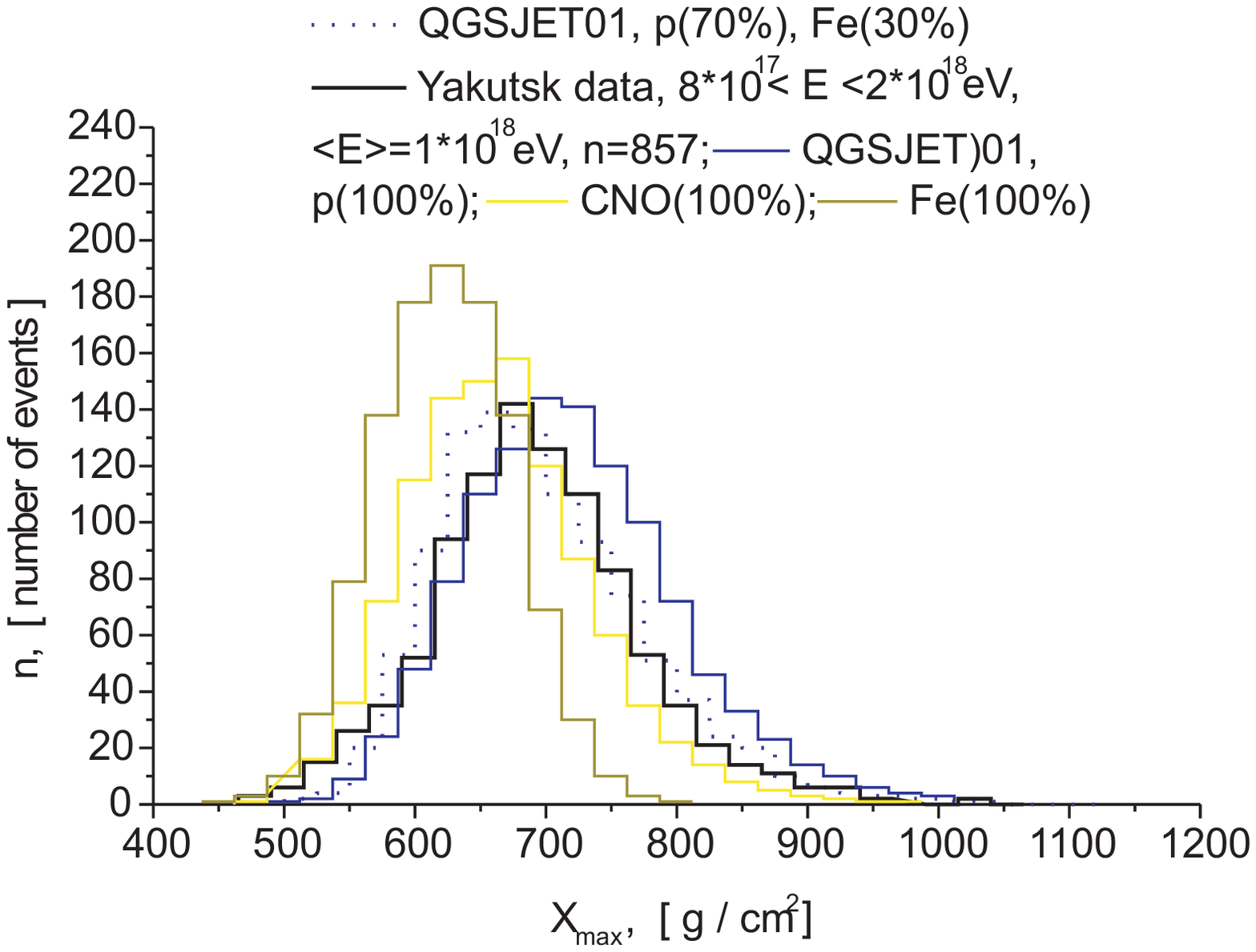}
\includegraphics[width=7.5cm]{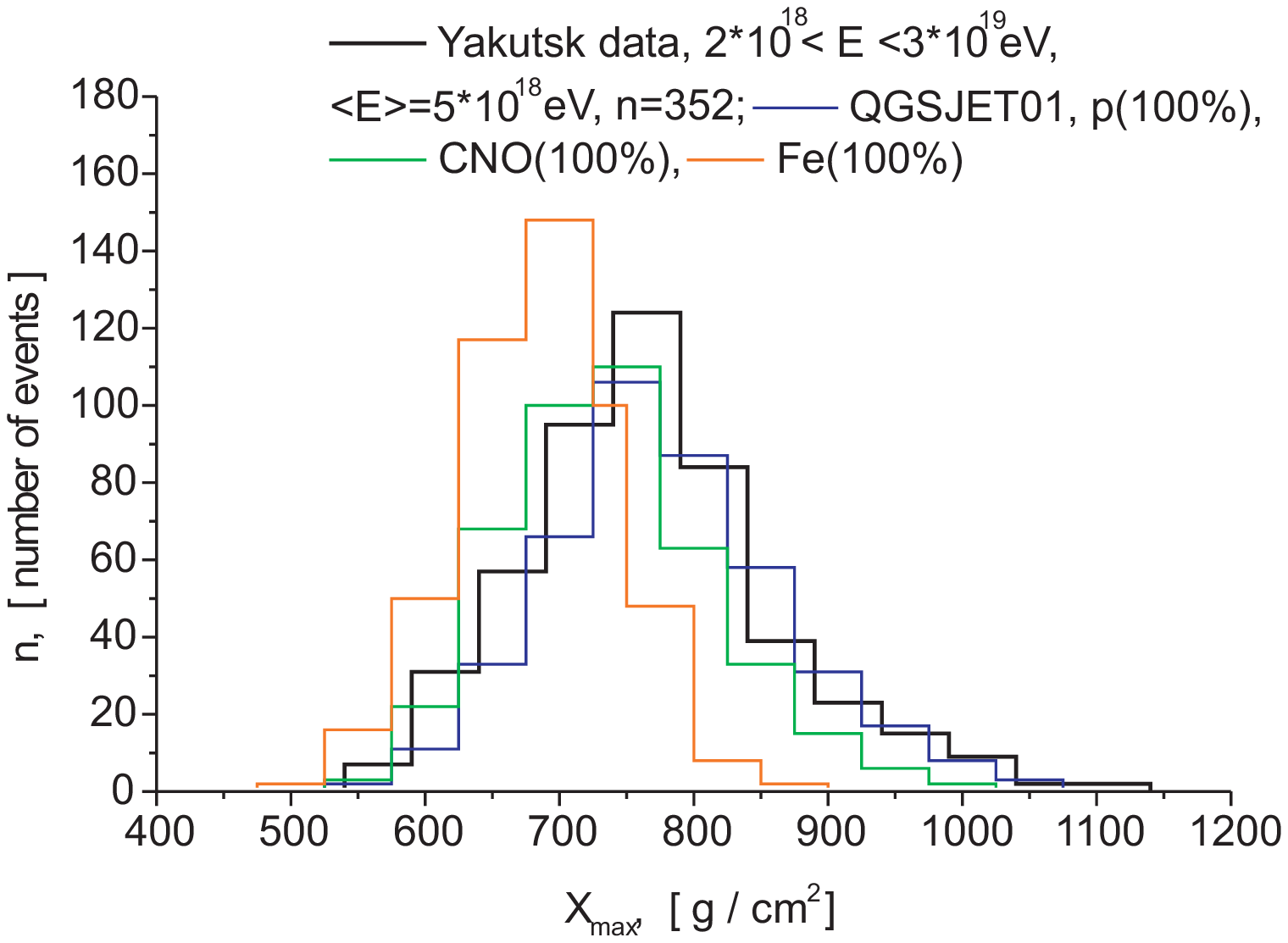}
\caption{Depth of maximum distributions for difference fixes energy.}
\label{fig3}
\end{figure}

\section{$\Xmax$ FLUCTUATIONS}

A large number of \^{C}erenkov detectors operating in the individual EAS events and also
the use of a new version of the QGSJET model allow to obtain quantitative estimations of
mass composition of PCR. For this aim we have compared experimental data (see
Fig.\ref{fig3}) and theoretical predictions by the QGSJET model for different primary
nucleus with the use of $\chis$ criterion. The value of $\chis$ has been determined by the
equality
\begin{equation}
\chis(\Xmax) = \Sigma \frac{\left[\Nexp(\Xmax) - \Nteor(\Xmax)\right]^2}{\Nteor(\Xmax)}
\label{eq2}
\end{equation}
where $\Nexp(\Xmax)$ is the experimental number of showers in the $\Delta \Xmax$ interval.
$\Nteor(\Xmax, A_{i})$ is the analogous number of showers calculated under the assumption
that the mass number of the nucleus is equal to $A_{i}$, and $P(A_{i})$ is the probability
of the fact that the shower with the energy $\E$ is formed by a primary particle $A_{i}$.
Then:
\begin{equation}
\Nteor(\Xmax) = \sum_{i = 1}^{n} P(A_{i}) \cdot \Nteor(\Xmax, A_{i})
\label{eq3}
\end{equation}

The analysis of form of the experimental distribution of $\Xmax$ at optimal value of
$\chis$ with a definite portion of probability doesn’t contradict to the following
relationship for five nuclei component:
\begin{description}
\item[$\bar \E = 5 \cdot 10^{17}$~eV:] p: $(39 \pm 11)$\%, $\alpha$: $(31 \pm 13)$\%, M:
$(18 \pm 10)$\%, H: $(7 \pm 6)$\%, Fe: $(5 \pm 4)$\%;
\item[$\bar \E = 1 \cdot 10^{18}$~eV:] p: $(41 \pm 8)$\%, $\alpha$: $(32 \pm 11)$\%, M:
$(16 \pm 9)$\%, H: $(6 \pm 4)$\%, Fe: $(5 \pm 3)$\% ;
\item[$\bar \E = 5 \cdot 10^{18}$~eV:] p: $(60 \pm 14)$\%, $\alpha$: $(21 \pm 13)$\%, M:
$(10 \pm 8)$\%, H: $(5 \pm 4)$\%, Fe: $(3 \pm 3)$\%.
\end{description}

Thus in the framework of the QGSJET model one can suppose that the mass composition of PCR
changes transferring from the energy range $(5 - 30) \cdot 10^{17}$~eV to energy range $(5
-30) \cdot 10^{18}$~eV. At $\E \ge 3 \cdot 10^{18}$~eV the primary cosmic radiation
consists from $\sim70$\% protons and helium nuclei, a portion of the rest nuclei in the
range where there is the second irregularity in the energetic spectrum type ``ankle''
doesn’t exceed $\sim30$\%. A high content of proton and helium nuclei in PCR in the
region of formation of ``ankle'' is most likely connected with an appreciable contribution
into the overall flux of cosmic radiation in the Earth’s vicinity of radiation coming from
beyond our Galaxy limits.

\section*{Acknowledgements}
This work has been financially supported by RFBR, grant \textnumero02--02--16380, grant
\textnumero03--02--17160 and grant INTAS \textnumero03-51-5112.


\begin{thebibliography}{9}
\bibitem{bib1}Knurenko S.P. et al. // Proc. 26th ICRC. Salt Lake Sity. V.1. P.372.
\bibitem{bib2}Song C., Cao Z., Dawson B.R. et al. Astroparticle Phys. 14 (2000) 7--13.
\bibitem{bib3}Sleptsova V.R. et al. // Nucl. Phys. B (Proc. Suppl.). 2003.V.122. P. 255.
\bibitem{bib4}K. Shinozaki et al. // Proc. 28th ICRC. Tsukuba. Japan. 2003. V. 3. P. 401.
\end{thebibliography}
\end{document}